\documentclass[pra,twocolumn,floatfix,a4paper,superscriptaddress]{revtex4}
\usepackage{bm,color,graphicx,amsmath,txfonts}
\usepackage{here}
\usepackage{array}
\usepackage{graphicx}
\usepackage[colorlinks=true,
citecolor=blue,
linkcolor=blue,
urlcolor=blue]{hyperref}
\usepackage{braket}
\usepackage{dsfont}
\usepackage[left=	2cm,top=2cm,right=1.2cm,bottom=2cm]{geometry}
\usepackage{tikz}
\usetikzlibrary{decorations.pathmorphing, positioning}

\usepackage{tikz}
\everymath{\displaystyle}

\makeatletter
\renewcommand{\fnum@figure}{\textbf{Fig.~\thefigure:}}
\makeatother

\makeatletter
\renewcommand{\@makecaption}[2]{%
	\textbf{#1} #2\par
}
\makeatother
\makeatletter
\renewcommand{\fnum@figure}{\normalsize\textbf{Fig.~\thefigure:}}
\makeatother
\makeatletter
\renewcommand{\@makecaption}[2]{%
	\begin{flushleft}
		\textbf{#1} #2
	\end{flushleft}
}
\makeatother

\begin{document}
\makeatletter
\renewcommand{\@biblabel}[1]{%
	\makebox[2.1em][l]{\fontsize{10}{13}\selectfont[#1]}}
\makeatother

\title{Multiparameter Quantum Metrology in Molecular Dimers}

\author{Omar Bachain}
\address{LPHE-Modeling and Simulation, Faculty of Sciences, Mohammed V University in Rabat, Rabat, Morocco}

\author{Mohamed \surname{Amazioug} }
\email{m.amazioug@uiz.ac.ma}
\address{LPTHE-Department of Physics, Faculty of Sciences, Ibnou Zohr University, Agadir 80000, Morocco}

\author{Rachid Ahl Laamara}
\address{LPHE-Modeling and Simulation, Faculty of Sciences, Mohammed V University in Rabat, Rabat, Morocco}
\address{Centre of Physics and Mathematics, CPM, Faculty of Sciences, Mohammed V University in Rabat, Rabat, Morocco}

\date{\today}

\begin{abstract}
	We investigate multiparameter quantum estimation in a molecular dimer composed of two dipole--dipole interacting two-level systems, focusing on the simultaneous estimation of temperature $T$ and detuning $\Lambda$. By employing a vectorization approach to derive the quantum Fisher information matrix, we analyze the precision limits of both simultaneous and individual estimation strategies. We show that simultaneous estimation outperforms the individual one in the near-resonant and low-temperature regime, where quantum coherence is enhanced, while its advantage is progressively reduced under detuned conditions and increasing temperature. Our results demonstrate that temperature acts as a key control parameter governing both estimation precision providing a unified perspective on quantum metrology. These findings highlight the potential of molecular quantum systems as realistic platforms for multiparameter quantum sensing.
\end{abstract}

\maketitle
\section{Introduction}

Quantum metrology aims to exploit uniquely quantum resources, such as coherence and entanglement, to enhance the precision of parameter estimation beyond classical limits~\cite{Giovannetti2006,Giovannetti2011,Paris2009}. In recent years, the extension of quantum estimation theory to the multiparameter scenario has attracted considerable attention, motivated by its relevance in quantum sensing, imaging, and thermodynamics~\cite{Helstrom1976,Holevo2011,Ragy2016}. In this context, the simultaneous estimation of multiple parameters may offer advantages over independent strategies, although such improvements are fundamentally constrained by the compatibility of the corresponding quantum measurements~\cite{Matsumoto2002}.

A central tool in quantum estimation theory is the quantum Fisher information matrix (QFIM), which determines the ultimate precision bounds through the quantum Cramér--Rao inequality~\cite{Braunstein1994,Paris2009}. The QFIM provides a fundamental characterization of the statistical distinguishability of quantum states and plays a central role in identifying optimal estimation strategies. In multiparameter quantum metrology, the interplay between quantum correlations, parameter compatibility, and measurement optimization remains an active area of research~\cite{Pezze2018,Albarelli2020}.

Recent theoretical and experimental advances have demonstrated that multiparameter quantum estimation can significantly improve the performance of quantum sensors in a variety of physical platforms, including atomic systems, superconducting circuits, optical interferometers, and solid-state devices. Nevertheless, identifying realistic quantum systems that simultaneously provide high estimation precision, analytical tractability, and experimental feasibility remains an important challenge.

Molecular quantum systems, such as dipole--dipole coupled two-level emitters, constitute promising candidates for quantum metrology owing to their controllable coherent interactions and robustness under experimentally accessible conditions~\cite{Ficek2005,Reina2014}. The interplay between dipole--dipole coupling, thermal fluctuations, and external control parameters gives rise to rich quantum dynamics that can be exploited for precision sensing and quantum technologies.

Despite considerable progress in quantum thermometry and multiparameter estimation, the simultaneous estimation of temperature and detuning in molecular dimers has received relatively little attention. Previous investigations have mainly focused on single-parameter estimation or on the characterization of quantum correlations separately. A comprehensive analytical study of the joint estimation of these physically relevant parameters within a realistic molecular dimer model is still lacking.

In this work, we investigate multiparameter quantum estimation in a molecular dimer composed of two dipole--dipole coupled two-level systems. We focus on the simultaneous estimation of temperature and detuning, two key parameters governing the system dynamics. Using a vectorization approach, we derive analytical expressions for the quantum Fisher information matrix and the corresponding quantum Cramér--Rao bounds, allowing a direct comparison between simultaneous and individual estimation strategies.

Our analysis reveals how the estimation precision is affected by temperature, detuning, dipole--dipole interaction strength, and molecular dissipation. We identify parameter regimes in which simultaneous estimation outperforms independent estimation and analyze the influence of the system parameters on the attainable precision limits. These results provide useful guidelines for optimizing multiparameter estimation protocols in realistic molecular quantum systems.

The paper is organized as follows. Section~\ref{sec2} introduces the molecular dimer model and derives its steady-state density matrix. Section~\ref{sec3} presents the analytical derivation of the quantum Fisher information matrix and discusses the corresponding multiparameter estimation strategies. Section~\ref{sec4} analyzes the numerical results and compares simultaneous and individual estimation protocols. Finally, Section~\ref{sec6} summarizes our main conclusions.
\section{Model and thermal state}\label{sec2}

\begin{figure}
	\centering
	\includegraphics[width=0.97\linewidth]{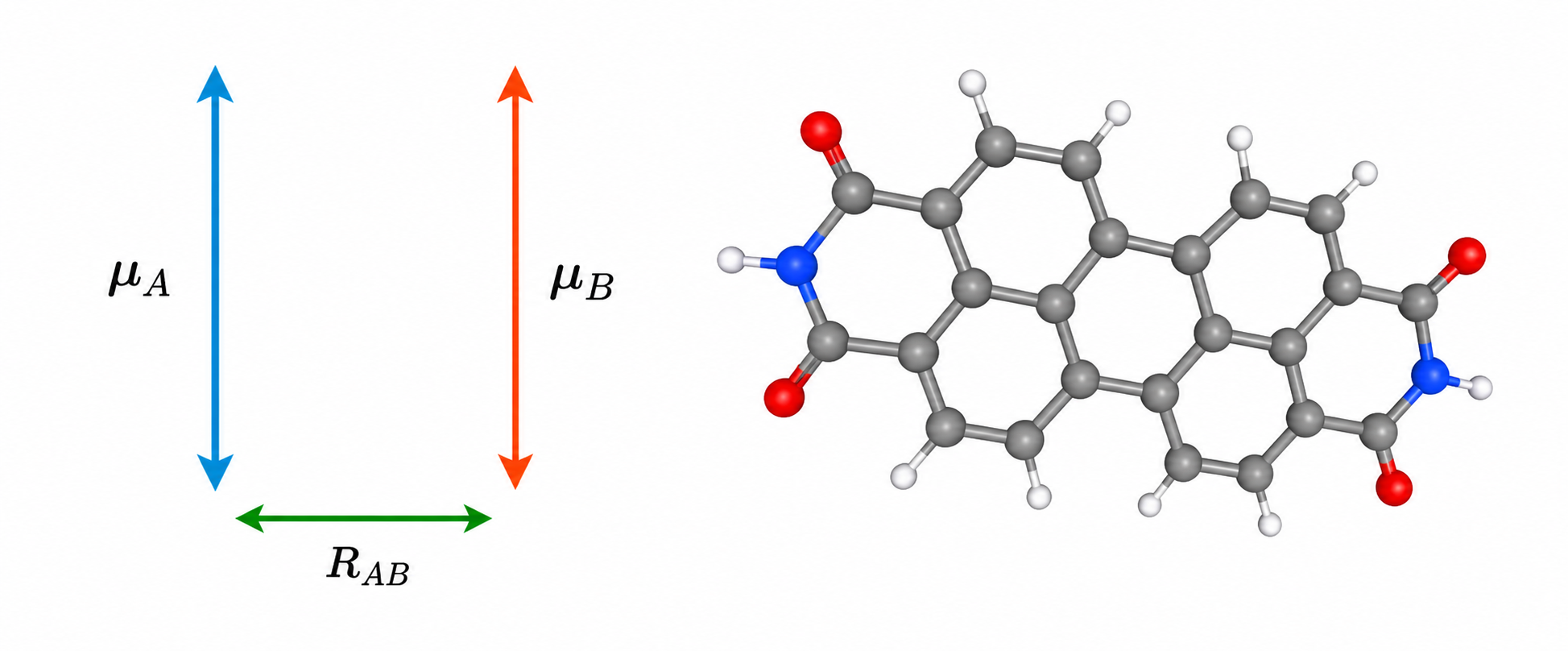}
	\caption{Schematic illustration of a molecular dimer system. The transition dipole moments $\mu_A$ and $\mu_B$ are oriented along parallel directions, while $R_{AB}$ denotes the intermolecular separation vector. On the right, the corresponding molecular structure is represented using a ball-and-stick model.}
	\label{fig1}
\end{figure}

We consider a molecular dimer composed of two interacting chromophores, labeled $A$ and $B$. Each chromophore is modeled as an effective two-level system corresponding to its electronic ground state and its first optically active excitonic excited state, an approximation widely adopted for molecular dimers exhibiting coherent excitonic interactions. The two chromophores are characterized by transition frequencies $\nu_A$ and $\nu_B$ and transition dipole moments $\mu_A$ and $\mu_B$. They are separated by an intermolecular vector $R_{AB}$ and interact through dipole--dipole coupling, as schematically illustrated in Fig.~\ref{fig1}. Such molecular dimers constitute realistic and experimentally accessible platforms for investigating coherent quantum dynamics, excitonic energy transfer, and quantum metrological protocols under experimentally relevant conditions \cite{Ficek2005,Reina2014,Reina2018}.

In the rotating-wave approximation and setting $\hbar = 1$, the total Hamiltonian of the system can be written as
\begin{equation}
	\mathcal{H} = -\frac{1}{2}\left( \nu_A \sigma_z^{(A)} + \nu_B \sigma_z^{(B)} \right)
	+ \frac{V_{AB}}{2} \left( \sigma_x^{(A)} \sigma_x^{(B)} + \sigma_y^{(A)} \sigma_y^{(B)} \right),
\end{equation}
where $V_{AB}$ denotes the dipole--dipole interaction strength. It is convenient to introduce the mean transition frequency and detuning as
\begin{equation}
	\nu_0 = \frac{\nu_A + \nu_B}{2}, \qquad \Lambda = \nu_A - \nu_B.
\end{equation}

In the computational basis $\{|00\rangle, |01\rangle, |10\rangle, |11\rangle\}$, the Hamiltonian can be expressed explicitly in operator form as
\begin{align}
	\mathcal{H}  &= -\nu_0 \, |00\rangle\langle 00| + \nu_0 \, |11\rangle\langle 11| \nonumber \\
	&\quad -\frac{\Lambda }{2} \, |01\rangle\langle 01|
	+ \frac{\Lambda }{2} \, |10\rangle\langle 10| \nonumber \\
	&\quad + V_{AB} \left( |01\rangle\langle 10| + |10\rangle\langle 01| \right).
\end{align}

This form shows that the states $|00\rangle$ and $|11\rangle$ are eigenstates of the Hamiltonian, while the single-excitation subspace $\{|01\rangle, |10\rangle\}$ is coupled through the interaction $V_{AB}$. Diagonalization of this subspace yields the dressed energies
\begin{equation}
	\mathcal{N} = \sqrt{\Lambda ^2 + 4 V_{AB}^2},
\end{equation}
leading to the eigenvalues
\begin{equation}
	\varepsilon_1 = -\nu_0, \quad \varepsilon_4 = \nu_0, \quad \varepsilon_{2,3} = \mp \frac{	\mathcal{N} }{2}.
\end{equation}
The corresponding eigenstates in the single-excitation manifold are coherent superpositions of $|01\rangle$ and $|10\rangle$. In the resonant case $\Lambda  = 0$, these reduce to maximally entangled Bell states, highlighting the role of dipole coupling in generating quantum correlations.

We assume that the system is initially prepared in a global thermal equilibrium state described by the Gibbs density operator
\begin{equation}
\varrho_{\mathrm{th}} = \frac{e^{-\mathcal{H} \,/T}}{Z},
\end{equation}
where $T$ denotes the temperature (in energy units) and the partition function is given by
\begin{equation}
	Z = 2\left[\cosh\left(\frac{\nu_0}{T}\right) + \cosh\left(\frac{	\mathcal{N} }{2T}\right)\right].
\end{equation}
This thermal state corresponds to the physically relevant situation in which the interacting molecular dimer is coupled to a common thermal environment and reaches global thermal equilibrium as a whole \cite{Binder2015,Allahverdyan2004,Elghaayda2026}. Therefore, the Gibbs state is not introduced merely to generate temperature-dependent populations, but represents the equilibrium state from which the subsequent multiparameter metrological analysis is performed. This equilibrium assumption is well justified for molecular dimers interacting with a common phononic or electromagnetic environment. In the computational basis, the density operator can be written explicitly as
\begin{align}
	\varrho_{\mathrm{th}} =&\;
	\varrho_{00} |00\rangle\langle 00|
	+ \varrho_{01} |01\rangle\langle 01|
	+ \varrho_{10} |10\rangle\langle 10|
	+ \varrho_{11} |11\rangle\langle 11| \nonumber \\
	&+ \varrho_c \, |01\rangle\langle 10|
	+ \varrho_c^{*} \, |10\rangle\langle 01|.
\end{align}

The populations are given by
\begin{align}
	\varrho_{00} &= \frac{1}{Z} e^{\nu_0/T}, \\
	\varrho_{11} &= \frac{1}{Z} e^{-\nu_0/T},
\end{align}
\begin{align}
	\varrho_{01} &= \frac{1}{Z	\mathcal{N} }
	\left[ \Lambda  \sinh\left(\frac{	\mathcal{N} }{2T}\right)
	+ 	\mathcal{N}  \cosh\left(\frac{	\mathcal{N} }{2T}\right) \right],
\end{align}
\begin{align}
	\varrho_{10} &= \frac{1}{Z	\mathcal{N} }
	\left[ -\Lambda  \sinh\left(\frac{	\mathcal{N} }{2T}\right)
	+ 	\mathcal{N}  \cosh\left(\frac{	\mathcal{N} }{2T}\right) \right],
\end{align}
while the coherence term reads
\begin{equation}
	\varrho_c = -\frac{2 V_{AB}}{	\mathcal{N}  Z}
	\sinh\left(\frac{	\mathcal{N} }{2T}\right).
\end{equation}

This expression reveals that quantum coherence naturally emerges in the single-excitation subspace due to the dipole--dipole interaction. In particular, $\varrho_c = 0$ when $V_{AB}=0$, indicating that coherence is entirely interaction-induced. Moreover, the competition between thermal fluctuations and coherent coupling governs the structure of the state: increasing temperature tends to suppress coherence, while finite detuning $\Lambda $ redistributes populations and breaks spectral symmetry.

The Gibbs state $\varrho_{\mathrm{th}}$ serves as the reference initial state for the quantum battery, representing a passive configuration from which no work can be extracted through unitary operations \cite{Lenard1978,Pusz1978}. Its structure encodes both thermal populations and interaction-induced coherences, which will play a central role in the charging dynamics analyzed in the following.

\section{Quantum Fisher information matrix}\label{sec3}
\label{secIII}

In this section, we develop the framework of multiparameter quantum estimation and derive an analytical expression for the quantum Fisher information matrix (QFIM) using a vectorization formalism. This approach circumvents the explicit diagonalization of the density matrix and provides an efficient computational scheme for finite-dimensional quantum systems \cite{vsafranek2018simple,Matsumoto2002,crowley2014tradeoff}.
Let ${H}$ be an $n$-dimensional Hilbert space and $\mathcal{B}({H})$ the space of linear operators acting on it. For any operator $A \in \mathcal{B}({H})$, its vectorized form is defined as \cite{gilchrist2009vectorization}
\begin{equation}
	|A\rangle\rangle = \sum_{k,l} a_{kl} \, |k\rangle \otimes |l\rangle .
\end{equation}
For a matrix
\begin{equation}
	A =
	\begin{pmatrix}
		a_{11} & a_{12} & \cdots & a_{1n} \\
		a_{21} & a_{22} & \cdots & a_{2n} \\
		\vdots & \vdots & \ddots & \vdots \\
		a_{n1} & a_{n2} & \cdots & a_{nn}
	\end{pmatrix},
\end{equation}
the vectorization corresponds to stacking its columns into a single vector,
\begin{equation}
	|A\rangle\rangle = (a_{11}, a_{21}, \ldots, a_{n1}, a_{12}, \ldots, a_{nn})^T.
\end{equation}

Equivalently,
\begin{equation}
	|A\rangle\rangle = (I \otimes A)\sum_{i=1}^{n} |i\rangle \otimes |i\rangle.
\end{equation}

The vectorization formalism satisfies the following identities:
\begin{align}
	(A \otimes B)|C\rangle\rangle &= |ACB^{T}\rangle\rangle, \\
	\langle\langle A|B\rangle\rangle &= \mathrm{Tr}(A^\dagger B), \\
	|A\rangle\rangle &= \sum_{k,l} a_{kl} |k\rangle |l\rangle .
\end{align}

We consider a parametric density operator $\varrho(\boldsymbol{\lambda})$ depending smoothly on a set of parameters
\[
\boldsymbol{\lambda}=(\lambda_1,\ldots,\lambda_m).
\]
The covariance matrix of any unbiased estimator is bounded by the quantum Cramér--Rao inequality \cite{Paris2009}
\begin{equation}
	\mathrm{Cov}(\hat{\boldsymbol{\lambda}})\ge \mathcal{K}^{-1}.
\end{equation}

The QFIM elements are defined as
\begin{equation}
	\mathcal{K}_{\mu\nu}
	=
	\frac{1}{2}
	\mathrm{Tr}\!\left[
	\varrho(\mathcal{D}_\mu\mathcal{D}_\nu+\mathcal{D}_\nu\mathcal{D}_\mu)
	\right],
\end{equation}
where the symmetric logarithmic derivatives (SLDs) satisfy
\begin{equation}
	\partial_{\lambda_\mu}\varrho
	=
	\frac{1}{2}(\mathcal{D}_\mu\varrho+\varrho\mathcal{D}_\mu).
\end{equation}

For a single parameter $\lambda$ \cite{Helstrom1976,Holevo2011},
\begin{equation}
	(\Delta\lambda)^2 \ge \frac{1}{\mathcal{K}(\lambda)}, \quad 
	\mathcal{K}(\lambda)=\mathrm{Tr}[\varrho \mathcal{D}_\lambda^2].
\end{equation}

Using the spectral decomposition of $\varrho$ \cite{Banchi2014,Sommers2003}, one obtains
\begin{equation}
	\mathcal{K}_{\mu\nu}
	=
	2\sum_{r_k+r_l>0}
	\frac{\langle k|\partial_{\lambda_\mu}\varrho|l\rangle
		\langle l|\partial_{\lambda_\nu}\varrho|k\rangle}{r_k+r_l}.
\end{equation}

To avoid diagonalization, we introduce the operator \cite{vsafranek2018simple}
\begin{equation}
	\eta=\varrho^T\otimes I + I\otimes\varrho,
\end{equation}
which allows one to rewrite the QFIM as
\begin{equation}
	\mathcal{K}_{\mu\nu}
	=
	2\,|\partial_{\lambda_\mu}\varrho\rangle\rangle^T
	\eta^{-1}
	|\partial_{\lambda_\nu}\varrho\rangle\rangle.
\end{equation}

The corresponding SLD operators are given by
\begin{equation}
	|D_\mu\rangle\rangle = 2\eta^{-1}|\partial_{\lambda_\mu}\varrho\rangle\rangle.
\end{equation}

For the two parameters $T$ and $\Lambda$, the vectorized derivatives read
\begin{align}\nonumber
	|\partial_T \varrho\rangle\rangle=
	\Big(\partial_T \varrho_{00} ,0,0,0,0,\partial_T\varrho_{01},\partial_T\varrho_{c},0,0,\,\partial_T\varrho_{c},\\
	\partial_T\varrho_{10},\,0,0,0,0,\partial_T\varrho_{11}\Big),\\\nonumber
	|\partial_\Lambda \varrho\rangle\rangle=
	\Big(\partial_\Lambda \varrho_{00} ,0,0,0,0,\partial_\Lambda\varrho_{01},\partial_\Lambda\varrho_{c},0,0,\,\partial_\Lambda\varrho_{c},\\
\partial_\Lambda\varrho_{10},\,0,0,0,0,\partial_\Lambda\varrho_{11}\Big).
\end{align}

The QFIM takes the matrix form
\begin{align*}
	\mathcal{K}=
	\begin{pmatrix}
		\mathcal{K}_{TT} & \mathcal{K}_{T\Lambda} \\
		\mathcal{K}_{ T\Lambda} & \mathcal{K}_{\Lambda\Lambda}
	\end{pmatrix},
\end{align*}
with elements
\begin{align}\nonumber
	\mathcal{K}_{TT}
	&=
\frac{
	(\mathcal{N} - 2\nu_0 )^2 \operatorname{sech}^2\!\left(\frac{\mathcal{N} - 2\nu_0 }{4T}\right)
	+
	(\mathcal{N} + 2\nu_0 )^2 \operatorname{sech}^2\!\left(\frac{\mathcal{N} + 2\nu_0 }{4T}\right)
}{
	16 T^4
},\\\nonumber
	\mathcal{K}_{T\Lambda}
	&=
	\frac{-\Lambda}{4 T^3 \mathcal{N}}
	\frac{
		\mathcal{N}
		+ \mathcal{N}\cosh\!\left(\frac{\mathcal{N}}{2T}\right)\cosh\!\left(\frac{\nu_0 }{T}\right)
		- 2\nu_0 \sinh\!\left(\frac{\mathcal{N}}{2T}\right)\sinh\!\left(\frac{\nu_0 }{T}\right)
	}{
		\left(
		\cosh\!\left(\frac{\mathcal{N}}{2T}\right)
		+ \cosh\!\left(\frac{\nu_0 }{T}\right)
		\right)^2
	},
\\\nonumber
	\mathcal{K}_{\Lambda \Lambda}
	&=	\frac{\cosh\!\left(\frac{\nu_0 }{T}\right)}{
		8 T^2 \mathcal{N}^4
		\left(\cosh\!\left(\frac{\mathcal{N}}{2T}\right)
		+ \cosh\!\left(\frac{\nu_0 }{T}\right)\right)^3
	}
		\Bigg[
		\left(32 T^2 V_{AB}^2 + \Lambda^2 \mathcal{N}^2\right)\\
		&\quad\times\cosh\!\left(\frac{\mathcal{N}}{T}\right)
		- \left(32 T^2 V_{AB}^2 - 3 \Lambda^2 \mathcal{N}^2\right)
		\Bigg]
	.
\end{align}

The quantum Cramér--Rao bounds read \cite{Prussing1986}
\begin{equation}
	(\Delta T)^2 \ge \frac{\mathcal{K}_{\Lambda\Lambda}}{\det\mathcal{K}}, \quad
	(\Delta \Lambda)^2 \ge \frac{\mathcal{K}_{TT}}{\det\mathcal{K}}.
\end{equation}

The corresponding minimal variances for the simultaneous estimation are

\begin{equation}
		(\Delta T)^2_{\text{sim}} =\frac{\mathcal{K}_{\Lambda\Lambda}}{\det\mathcal{K}}, \qquad(\Delta \Lambda)^2_{\text{sim}} =\frac{\mathcal{K}_{TT}}{\det\mathcal{K}}.
\end{equation}

We now turn to the scenario of individual parameter estimation. In this approach, the parameters are assumed to be statistically independent, such that estimating one parameter does not influence the precision of the others. This condition is satisfied when the off-diagonal elements of the quantum Fisher information matrix vanish, i.e., $\mathcal{K}_{ij} = 0$ for $i \neq j$. Under this assumption, the estimation problem reduces to independent single-parameter estimation tasks. This implies

\begin{figure*}[t!]
	\centering
	\includegraphics[width=0.45\linewidth]{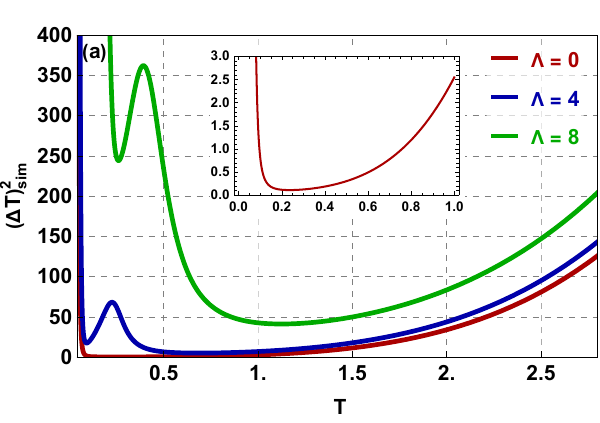}\hspace*{0.3cm}
	\includegraphics[width=0.45\linewidth]{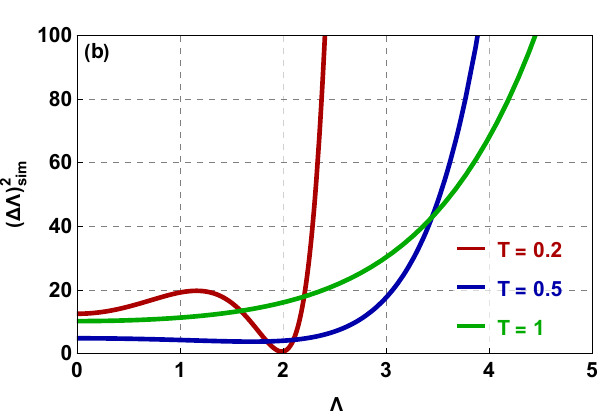}\\
	\includegraphics[width=0.45\linewidth]{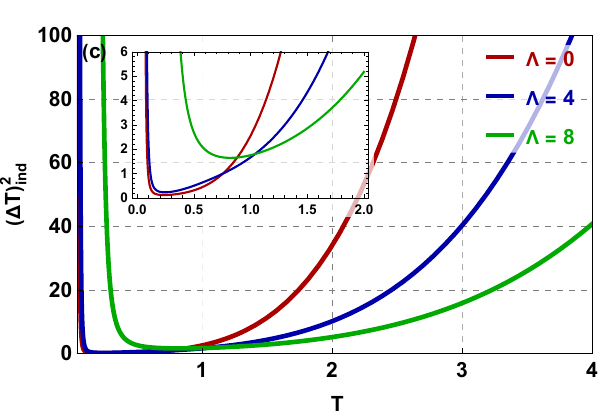}\hspace*{0.3cm}
	\includegraphics[width=0.45\linewidth]{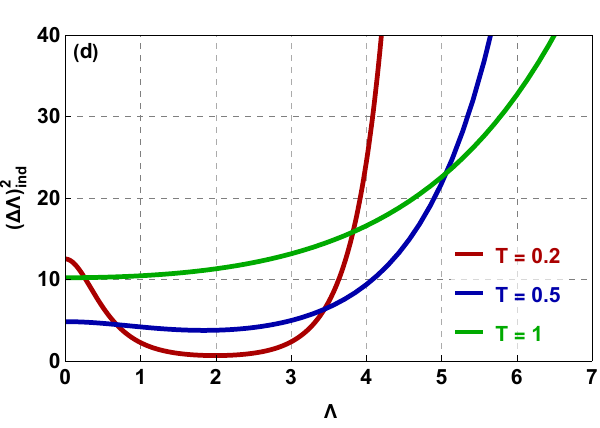}
	\caption{
		Estimation variances associated with temperature $T$ and detuning $\Lambda$ for simultaneous and individual estimation strategies. The upper row corresponds to simultaneous estimation, while the lower row refers to individual estimation. The left column shows $(\Delta T)^2$ as a function of temperature for different values of $\Lambda$, and the right column shows $(\Delta \Lambda)^2$ as a function of $\Lambda$ for different temperatures. Results are obtained for $\nu_0 = 1$ and $V_{AB} = 0.1$.
	}
	\label{fig2}
\end{figure*}
\begin{equation}
	(\Delta T)^2_{\text{ind}} \geq\frac{1}{\mathcal{K}_{TT}}, \qquad(\Delta \Lambda)^2_{\text{ind}} \geq\frac{1}{\mathcal{K}_{\Lambda\Lambda}}.
\end{equation}
Saturating the above inequalities, one obtains
\begin{equation}
	(\Delta T)^2_{\text{ind}} =\frac{1}{\mathcal{K}_{TT}}, \qquad(\Delta \Lambda)^2_{\text{ind}} =\frac{1}{\mathcal{K}_{\Lambda\Lambda}}.
\end{equation}
We quantify the relative performance of the two estimation strategies through the ratio
\begin{equation}
	\Gamma(T,\Lambda)=\frac{\Delta_{\mathrm{sim}}(T,\Lambda)}{\Delta_{\mathrm{ind}}(T,\Lambda)}
	=\frac{1}{2}\,
	\frac{(\Delta T)^2_{\text{sim}} +(\Delta \Lambda)^2_{\text{sim}} }
	{(\Delta T)^2_{\text{ind}} +(\Delta \Lambda)^2_{\text{ind}}}.
\end{equation}
A simultaneous estimation strategy outperforms the individual one when $\Gamma(T,\Lambda)<1$, i.e., $\Delta_{\mathrm{Sim}}<\Delta_{\mathrm{Ind}}$. Conversely, $\Gamma(T,\Lambda)>1$ (i.e., $\Delta_{\mathrm{sim}}>\Delta_{\mathrm{ind}}$) indicates that individual estimation yields higher precision.

The SLDs are given by
\begin{equation}
	 \mathcal{D}^T =
\begin{pmatrix}
	\mathcal{D}^T_{11} & 0 & 0 & 0 \\
	0 & \mathcal{D}^T_{22} & \frac{V_{AB}}{T^2} & 0 \\
	0 & \frac{V_{AB}}{T^2} & \mathcal{D}^T_{33} & 0 \\
	0 & 0 & 0 & \mathcal{D}^T_{44}
\end{pmatrix},
\end{equation}
with
\begin{align}\nonumber
	\mathcal{D}^T_{11} &= \frac{-2\nu e^{-\nu/T}-2\nu \cosh\!\left(\frac{\mathcal{N}}{2T}\right)+\mathcal{N}\sinh\!\left(\frac{\mathcal{N}}{2T}\right)}
	{2T^2\left(\cosh\!\left(\frac{\mathcal{N}}{2T}\right)+\cosh\!\left(\frac{\nu}{T}\right)\right)}, \\\nonumber
	\mathcal{D}^T_{22} &= \frac{-\Lambda\left(\cosh\!\left(\frac{\mathcal{N}}{2T}\right)+\cosh\!\left(\frac{\nu}{T}\right)\right)+\mathcal{N}\sinh\!\left(\frac{\mathcal{N}}{2T}\right)+2\nu\sinh\!\left(\frac{\nu}{T}\right)}
	{2T^2\left(\cosh\!\left(\frac{\mathcal{N}}{2T}\right)+\cosh\!\left(\frac{\nu}{T}\right)\right)}, \\\nonumber
	\mathcal{D}^T_{33} &= \frac{\Lambda\left(\cosh\!\left(\frac{\mathcal{N}}{2T}\right)+\cosh\!\left(\frac{\nu}{T}\right)\right)+\mathcal{N}\sinh\!\left(\frac{\mathcal{N}}{2T}\right)+2\nu\sinh\!\left(\frac{\nu}{T}\right)}
	{2T^2\left(\cosh\!\left(\frac{\mathcal{N}}{2T}\right)+\cosh\!\left(\frac{\nu}{T}\right)\right)}, \\
	\mathcal{D}^T_{44} &= \frac{2\nu e^{\nu/T}+2\nu \cosh\!\left(\frac{\mathcal{N}}{2T}\right)+\mathcal{N}\sinh\!\left(\frac{\mathcal{N}}{2T}\right)}
	{2T^2\left(\cosh\!\left(\frac{\mathcal{N}}{2T}\right)+\cosh\!\left(\frac{\nu}{T}\right)\right)},
\end{align}
and 
\begin{equation}
	\mathcal{D}^\Lambda =
	\begin{pmatrix}
		D_{11}^\Lambda & 0 & 0 & 0 \\
		0 & D_{22}^\Lambda & D_{23}^\Lambda & 0 \\
		0 & D_{23}^\Lambda & D_{33}^\Lambda & 0 \\
		0 & 0 & 0 & D_{44}^\Lambda
	\end{pmatrix},
\end{equation}
with
\begin{align}\nonumber
	D_{11} &= -\frac{\Lambda \sinh\!\left(\frac{\mathcal{N}}{2T}\right)}
	{2T\,\mathcal{N}\left(\cosh\!\left(\frac{\mathcal{N}}{2T}\right)+\cosh\!\left(\frac{\nu}{T}\right)\right)}, \\\nonumber
	D_{22} &= \frac{
		\Lambda^2 \mathcal{N}^3
		- \Lambda \mathcal{N}^5 \frac{\sinh\!\left(\frac{\mathcal{N}}{2T}\right)}
		{\cosh\!\left(\frac{\mathcal{N}}{2T}\right)+\cosh\!\left(\frac{\nu}{T}\right)}
		+ 8 T V_{AB}^2 \mathcal{N}^2 \tanh\!\left(\frac{\mathcal{N}}{2T}\right)
	}{
		2T\,\mathcal{N}^3(4V_{AB}^2+\Lambda^2)
	}, \\\nonumber
	D_{23} &= \frac{
		V_{AB}\Lambda\left(-\mathcal{N}^3 + 2T(4V_{AB}^2+\Lambda^2)\tanh\!\left(\frac{\mathcal{N}}{2T}\right)\right)
	}{
		T\,\mathcal{N}^3(4V_{AB}^2+\Lambda^2)
	}, \\\nonumber
	D_{33} &= \frac{
		-\Lambda^2 \mathcal{N}^3
		- \Lambda \mathcal{N}^5 \frac{\sinh\!\left(\frac{\mathcal{N}}{2T}\right)}
		{\cosh\!\left(\frac{\mathcal{N}}{2T}\right)+\cosh\!\left(\frac{\nu}{T}\right)}
		- 4 T V_{AB}^2 \mathcal{N}^2 \tanh\!\left(\frac{\mathcal{N}}{2T}\right)
	}{
		2T\,\mathcal{N}^3(4V_{AB}^2+\Lambda^2)
	}, \\
	D_{44} &= -\frac{\Lambda \sinh\!\left(\frac{\mathcal{N}}{2T}\right)}
	{2T\,\mathcal{N}\left(\cosh\!\left(\frac{\mathcal{N}}{2T}\right)+\cosh\!\left(\frac{\nu}{T}\right)\right)}.
\end{align}
Finally, the SLD operators associated with $T$ and $\Lambda$ are obtained explicitly. In general, these operators do not commute, which prevents the existence of a common eigenbasis. However, a weaker condition ensuring the asymptotic attainability of the multiparameter quantum Cramér--Rao bound is given by \cite{ragy2016compatibility,vrehavcek2018optimal}
\begin{equation}
	\mathrm{Tr}\big(\varrho\, [\mathcal{D}_T, \mathcal{D}_{\Lambda}]\big) = 0.
\end{equation}

This compatibility condition provides the fundamental criterion for assessing the asymptotic attainability of the multiparameter quantum Cramér--Rao bound. For the molecular dimer considered in the present work, we have verified that this weak compatibility condition is satisfied throughout the parameter regime investigated in our numerical analysis. Consequently, although the SLD operators do not generally commute, the simultaneous estimation of temperature and detuning remains asymptotically achievable within the considered operating regime. This result further supports the validity of the multiparameter estimation framework adopted in this work.

The behavior of the estimation variances reveals a clear dependence on both temperature and detuning. In the simultaneous estimation scenario, $(\Delta T)^2_{\mathrm{sim}}$ exhibits a pronounced minimum at finite temperature, indicating the existence of an optimal working point where thermal fluctuations and quantum correlations balance each other. As shown in Fig.~\ref{fig2}(a), this optimal temperature shifts towards higher values and the minimal variance increases as the detuning $\Lambda$ increases. 
In the resonant case ($\Lambda = 0$), the estimation precision is significantly enhanced, with a sharp minimum at low temperature. This behavior reflects the strong role of coherent interactions, which maximize quantum sensitivity. In contrast, under detuned conditions ($\Lambda \neq 0$), the curves become broader and the minimum shifts, indicating a degradation of estimation precision due to reduced effective coupling.
A similar trend is observed for $(\Delta \Lambda)^2_{\mathrm{sim}}$ in Fig.~\ref{fig2}(b), where the variance increases with detuning and temperature, and lower temperatures provide better estimation precision. Notably, the dependence on $\Lambda$ becomes more pronounced at higher temperatures, highlighting the interplay between thermal noise and parameter sensitivity.
For the individual estimation strategy, Figs.~\ref{fig2}(c) and \ref{fig2}(d) show that the variances are systematically larger compared to the simultaneous case. The optimal temperature still exists but is less pronounced, and the estimation precision deteriorates more rapidly with increasing detuning. This clearly demonstrates the advantage of the simultaneous estimation scheme.
Overall, the results indicate that the optimal estimation regime is achieved at low temperature and near-resonant conditions ($\Lambda \approx 0$), where quantum correlations are strongest. Away from resonance, the detuning suppresses coherent effects, leading to reduced sensitivity and higher estimation uncertainties.
\begin{figure}
	\centering
	\includegraphics[width=0.97\linewidth]{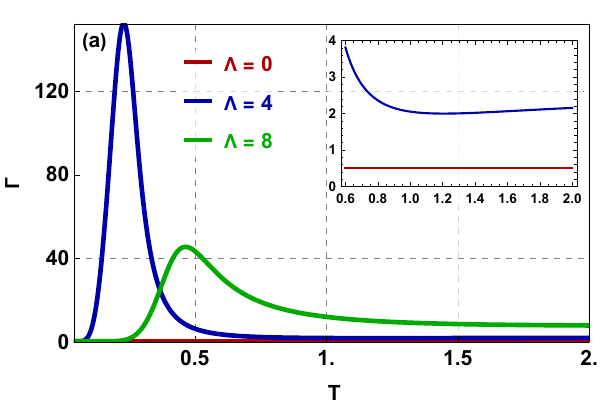}
	\includegraphics[width=0.97\linewidth]{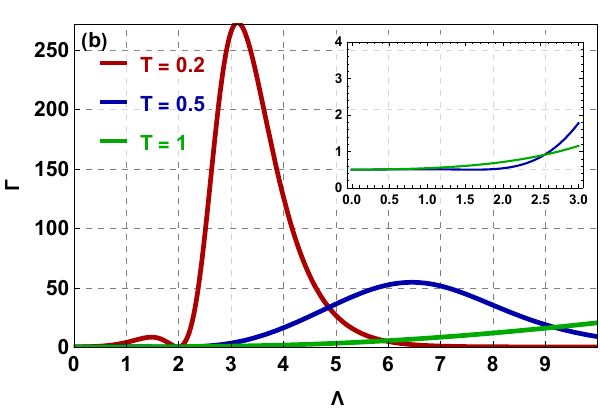}
\caption{
	Plot of the ratio $\Gamma$ of minimal total variances for the estimation of $T$ and $\Lambda$. (a) $\Gamma$ as a function of $T$ for different values of $\Lambda$. (b) $\Gamma$ as a function of $\Lambda$ for different temperatures. The other parameters are fixed at $\nu_0 = 1$ and $V_{AB} = 0.1$.
}
	\label{fig3}
\end{figure}

The ratio $\Gamma$ provides a direct quantitative comparison between simultaneous and individual estimation strategies. As shown in Fig.~\ref{fig3}, the condition $\Gamma < 1$ identifies the regime where simultaneous estimation outperforms the individual one, while $\Gamma > 1$ indicates the opposite behavior.
In the resonant case ($\Lambda = 0$), $\Gamma$ remains below unity over the whole temperature range, demonstrating a clear advantage of the simultaneous estimation strategy. This behavior reflects the constructive role of quantum correlations, which enhance the joint estimation precision when the system operates at resonance.
In contrast, under detuned conditions ($\Lambda \neq 0$), $\Gamma$ exhibits pronounced peaks as a function of temperature, as visible in Fig.~\ref{fig3}(a). These peaks correspond to regions where $\Gamma \gg 1$, indicating a strong degradation of the simultaneous estimation precision compared to the individual strategy. The position and amplitude of these peaks depend on the value of the detuning, shifting toward higher temperatures and becoming broader as $\Lambda$ increases.
A similar behavior is observed when analyzing $\Gamma$ as a function of $\Lambda$ in Fig.~\ref{fig3}(b). At low temperature, $\Gamma$ displays sharp maxima at intermediate detuning values, signaling a breakdown of the simultaneous estimation advantage. As the temperature increases, these peaks become smoother and shift, reflecting the interplay between thermal fluctuations and detuning-induced suppression of quantum coherence.
From a global perspective, the results show that simultaneous estimation is optimal in the near-resonant regime ($\Lambda \approx 0$) and at low temperature, where quantum correlations are strongest. Away from resonance or at higher temperatures, the advantage progressively disappears, and individual estimation can become more efficient.
\section{Conclusion}\label{sec6}

In this work, we have investigated the multiparameter quantum estimation properties of a molecular dimer system, focusing on the simultaneous estimation of the temperature $T$ and the detuning $\Lambda$. By employing a vectorization approach, we derived analytical expressions for the quantum Fisher information matrix and the corresponding quantum Cramér--Rao bounds, allowing a direct comparison between simultaneous and individual estimation strategies.

Our results show that the simultaneous estimation strategy can outperform individual estimation in appropriate parameter regimes, particularly at low temperatures and near resonance, where the system exhibits enhanced parameter sensitivity. In contrast, increasing the temperature or moving away from resonance progressively reduces the attainable precision because of thermal fluctuations and the reduced distinguishability of the quantum states.

We have also analyzed the influence of the dipole--dipole interaction, temperature, and detuning on the attainable precision limits. The interplay among these parameters determines the optimal operating regime of the molecular dimer and provides useful guidelines for enhancing multiparameter estimation performance. The analytical expressions obtained for the QFIM further provide a transparent description of how each physical parameter contributes to the estimation process.

Overall, our results demonstrate that molecular dimers constitute promising platforms for multiparameter quantum metrology. Beyond providing analytical precision bounds, the present work contributes to a deeper understanding of the simultaneous estimation of multiple physically relevant parameters in interacting quantum systems. We expect that the proposed framework can be extended to more complex molecular architectures, open quantum systems, and other solid-state platforms, thereby offering new opportunities for quantum sensing and quantum thermometry.

\end{document}